\documentclass[showpacs,twocolumn,preprintnumbers,amsmath,amssymb]{revtex4}
\usepackage{graphicx}% Include figure files

\usepackage{bm}% bold math
\begin{document}
\title{An Alternative Mechanism of Heat Transport in Icosahedral $i$-AlPdMn Quasicrystals}

\author{S.E. Krasavin}
\email{krasavin@thsun1.jinr.ru}
\affiliation{ Joint Institute
for Nuclear Research,
Bogoliubov Laboratory of Theoretical Physics\\
141980 Dubna, Moscow region, Russia}

\date{\today}

\begin{abstract}
We propose a new theoretical approach to analyze the
experimental data for thermal conductivity of single-grain
$i$-AlPdMn quasicrystals. The interpretation is based on the
picture where cluster interfaces are the main source of phonon
scattering at low temperatures. The scattering due to strain
fields of cluster interface is considered in terms of finite
dislocation wall or, equivalently, wedge disclination dipole.
Our numerical calculations show that experimentally observed
thermal conductivity in a wide temperature range can be well
fitted by a combination of wedge discliantion dipole scattering
and quasiumklapp scattering processes.
\end{abstract}


\maketitle

%\begin{center}
%{\bf I. INTRODUCTION}\\
%\end{center}
It is well established  experimentally  that icosahedral
quasicrystals appear to have a glass-like thermal conductivity
\cite{chern1,pohl}. Most of the early studies described the
low-temperature thermal conductivity (below 1 K) of
quasicrystals in the framework of the tunneling states model
(TLS) as well as amorphous solids \cite{chern1,chern2}. If the
appearance of the tunneling states in amorphous materials can be
explained by the presence of disorder, the physical nature of
their existence in quasicrystals is not yet understood.
Quasicrystals are long range ordered structures with a lack of
periodicity, so, the tunneling states then can be induced by
some particular type of disorder (e.g. phason disorder
\cite{birge}).
%However, tunneling states have been observed in acoustic measurements for phason-free
%single-grain Al-Cu-Fe quasicrystals\cite{bert} .
 At present, in some works  the low temperature thermal
conductivity of $i$-YMgZn and $i$-AlPdMn icosahedral
quasicrystals has been considered without tunneling states
\cite{ott1,ott2}. The good fit for the low and intermediate
temperatures was reached with the assumption that the total
relaxation rate is the combination of the Casimir, stacking
fault and quasiumklapp scattering processes.
%The Casimir
%relaxation rate is proportional to the average sound velocity
%and inversely to the phonon mean free path $l_{0}$ (average
%sample size). As well known, the low-temperature behaviour of
%thermal conductivity can be dependent prominently upon the value
%of $l_{0}$. In \cite{ott2} the value of $l_{0}$ has been
%extracted from the experimental data. To our mind, a set of
%different $l_{0}$, corresponding to the size of one sample (or
%identical samples with the same composition), should be used in
%calculations and compared with experimental results. This
%approach will have allowed to clarify how sensitive is the
%thermal conductivity  to the sample size for these compounds.

\hspace*{0.35cm} On the other hand, the presence of cluster
interfaces in  AlPdMn single quasicrystals that have been
recently observed experimentally in \cite{ebert} can affect
thermal conductivity as well. As is well known, a grain boundary
between two clusters can be described in the framework of the
disclination concept\cite{li1}. The far strain fields caused by
wedge disclination dipoles (WDD) are equivalent to those from
{\it finite} wall of edge dislocations with parallel Burgers
vectors \cite{li2}. Considering a grain boundary as an array of
edge dislocations in the standard approach \cite{lott},
WDD-based model allows to study the effect of the grain
boundaries finitness on the electric and thermal transport
\cite{kras3,kras1}.

\hspace*{0.35cm}  The contribution to thermal conductivity due
to the phonon scattering by static strain fields of WDD has been
estimated recently in \cite{kras1,osipov}. It was found that WDD
are very specific sources of scattering via their long-range
strain fields. A combination of WDD and Rayleigh type of
scattering processes allowed to explain the experimentally
observed thermal conductivity of some dielectric glasses in a
wide temperature range\cite{kras1}. We have found that only this
type of all possible dipoles should be involved in calculations
to fit the experimental data for thermal conductivity of
dielectric glasses. The effect of other disclination defects
(e.g. twist disclination dipoles, wedge disclination loop) on
thermal transport has been analyzed as well, and no glass-like
thermal conductivity was found (See e.g. \cite{krasa}).

In this Letter, we show that the observed in \cite{ott2}
temperature behaviour of the thermal conductivity of  $i$-AlPdMn
quasicrystals can be explained if the cluster grain boundary
phonon scattering (WDD scattering) combined with quasiumklapp
scattering  is included (without consideration of the Casimir
mechanism of scattering).

In our picture, WDD with separation $2L$ (cluster size) are
distributed in the $XY$ plane, and their lines are oriented
along the $Z$-axis. We consider WDD where the axes of rotation
are not shifted relative to disclination lines (biaxial WDD). If
the dipole arm is oriented along the $x$-axis, an effective
perturbation energy due to the strain field caused by a single
WDD is
\begin{eqnarray}
U(x,y)=\frac{\hbar qv\gamma \Omega (1-2\sigma)}{4\pi(1-\sigma)}\ln\frac{(x+L)^2+y^2}{(x-L)^2+y^2},
\end{eqnarray}
where $q$ is the phonon wavevector, $v$ is the sound velocity, $\gamma$
is the Gr\"{u}neisen constant, $\Omega$ is the axial vector (Frank vector) directed
along the disclination line and equal to $\pm 72^{o}$ for icosahedral quasicrystals\cite{nelson1},
$\sigma$ is the Poisson constant.

For the chosen geometry, in veiw of the Eq.(1) the problem of
scattering reduces to the two-dimensional case as for edge
dislocation \cite{gant}. Then, a mean free path arising due to
the phonon scattering by static strain fields of WDD within the
generally accepted deformation potential approach is given by
\cite{kras1}
\begin{eqnarray}
\l^{-1}_D(q) &=& 2A^2(\Omega L)^2n_{def}q\Bigl (J_0^2(2qL)+J_1^2(2qL) \nonumber \\
&-&\frac{1}{2qL}J_0(2qL)J_1(2qL)\Bigr ),
\end{eqnarray}
where $A=\gamma (1-2\sigma)/(1-\sigma)$,  $n_{def}$ is the arial
density of WDD, $J_{n}(t)$ are the Bessel functions. Notice also
that to get Eq.(2) we considered the elastic scattering of
phonons with wave vector $q$ within the Born approximation.

The total mean free path is written as
\begin{eqnarray}
\l(\omega ) &=&
\Bigl(l^{-1}_{D}(\omega)+l^{-1}_{qu}(\omega)\Bigr)^{-1}+l_{min},
\end{eqnarray}
with $l_{qu}$ being  the mean free path due to quasiumklapp
processes (See \cite{kalug}). To fit the experimentally observed
thermal conductivity \cite{chern1,ott2} for $i$-AlPdMn
quasicrystals, we use the weaker temperature dependence for
$l_{qu}$ than that proposed in \cite{kalug} where $l_{qu}\propto
\omega^2 T^4$. According to \cite{ott2} it takes the form
\begin{eqnarray}
\l^{-1}_{qu}(\omega )=B\frac{\omega^2}{v_s}T^2,
\end{eqnarray}
where $B$ is the fitting constant. The last term in Eq.(3)
describes the least possible mean free path of propagating
acoustic phonons. The experimental evidence for the $l_{min}$
introduction in Eq.(2) follows from inelastic neutron scattering
experiments in\cite{boiss}. It was found that unbroadened
acoustic modes can exist only for wave vectors $k\leq 0.35
\AA^{-1}$ that corresponds to $l_{min}\sim 18\AA $.

To calculate the temperature dependence of thermal conductivity
with the mean free path given by Eq.(3) we use the following
kinetic formula written in the dimensionless form
\begin{equation}
\kappa = \frac{k_B^4T^3}{2\pi^2\hbar^3v_s^2}
\int_0^{\Theta/T}x^4e^x(e^x-1)^{-2}l(x)dx,
\end{equation}
where $k_{B}$ and $\hbar$ are Boltzmann's and reduced Planck constant,
$\Theta $ is the Debye temperature, $x=\hbar\omega/k_BT$.

Fig.1 shows the experimental points for thermal conductivity over a wide
temperature range of two $i$-AlPdMn samples from\cite{chern1,ott2} together with
theoretical curves.
\begin{figure}[h]
\includegraphics*[width=9cm,height=7cm]{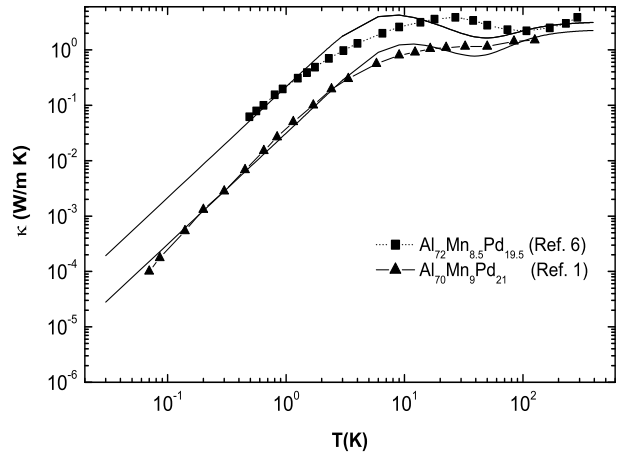}
\caption{Fig.1.  Quasilattice thermal conductivity vs
temperature for two $i$-AlPdMn samples from Refs.(1) and (6).
Solid lines are fits according to  Eqs.(3) and (5). The fitting
parameters for the upper experimental curve: $n_{def}=1.4\times
10^{10}$ cm$^{-2}$, $B=1.3\times 10^{-18}$ s K$^{-2}$,
$l_{min}=18\AA$, and for the lower curve $n_{def}=9\times
10^{10}$  cm$^{-2}$, $B=3.2\times 10^{-18}$ s K$^{-2}$,
$l_{min}=13\AA$. The common parameter set for both theoretical
curves is $\Theta =380$ K, $v_{s}=4\times 10^5$ cm c$^{-1}$,
$2L=20\AA $.   }
\end{figure}
It is seen from the plot, that there is a good agreement between
the  fitting curves and experimental data for $T$ below $1$ K
where $\kappa\sim T^2$. In our scheme, the main contribution to
$\kappa(T)$ at lowest $T$ is due to the phonon scattering by
cluster grain boundaries that we consider in the framework of
the biaxial WDD concept. One can obtain from Eq.(2), $l_{D}\sim
\omega^{-1}$ when $\lambda>2L$ that leads to the observed
$\kappa\sim T^2$ at low temperatures. It should be noted that
the best fit in our calculations for both experimental curves
was achieved when $2L$ and $l_{min}$ are equal approximately  to
$20\AA $.  This value  is in agreement with the average cluster
size found for $i$-AlPdMn single quasicrystals by scanning
tunneling microscopy in \cite{ebert}.

For $\lambda<2L$, $l_{D}\rightarrow const $, and the crossover
to $\kappa\sim T^3$ takes place at $T^{*}\simeq\hbar
v_{s}/2Lk_{B}=15\div 20$ K, if only the WDD source of scattering
(2) is present instead of (3). The formula for $T^{*}$ can be
derived from the condition $\lambda\sim 2L$ using the dominant
phonon approximation\cite{kras1}. This crossover, however, is
absent on the plot because of the dominanting quasiumklapp
processes at these temperatures. A qualitative fit between
calculated  and experimental $\kappa(T)$ is seen in Fig.1 in the
maxima region. This is the result of combination of two
scattering processes in calculations: scattering due to biaxial
WDD and quasiumklapp scattering. As noted in\cite{ott2} a
maximum in $\kappa(T)$ becomes more sharp if disorder decreases
(like for the upper experimental curve indicated by squares in
Fig.1). In the model maxima exist between 3 and 20 K where the
concentration of the defects (the source of disorder)
$n_{def}\sim 10^{10}$ cm$^{-2}$ for the upper experimental curve
and $n_{def}\sim 10^{11}$ cm$^{-2}$ for the lower one. When
$n_{def}>10^{11}$cm$^{-2}$ we found the variation of $\kappa $
as typical for amorphous solids (See the solid line in Fig.2)
with the plateau-like regime for $T$ between $10$ and $40$ K.

\begin{figure}[h]
\includegraphics*[width=9cm,height=7cm]{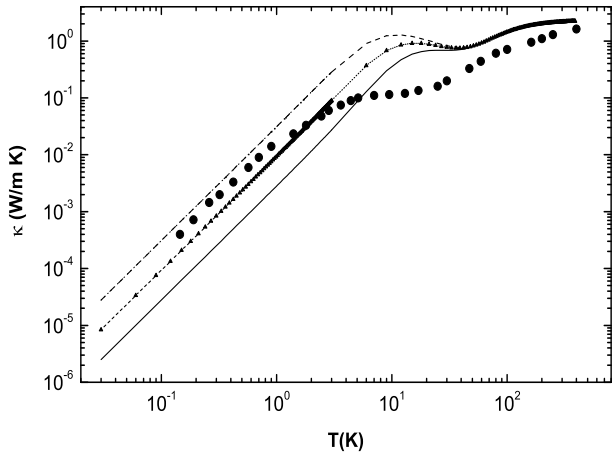}
\caption{Fig.2. Thermal conductivity vs temperature evaluated
with Eqs.(3) and (5) for concentration of WDD $n_{def}=9\times
10^{10}$ cm$^{-2}$ (dashed line), $n_{def}=3\times 10^{11}$
cm$^{-2}$ (dotted line), $n_{def}=9.7\times 10^{11}$ cm$^{-2}$
(solid line). The dots correspond to the experimental data for
SiO$_2$ from Ref.[19]. }
\end{figure}
In our previous work\cite{kras1} the good fit for amorphous
SiO$_{2}$ $\kappa (T)$ was obtained by combining the biaxial WDD
scatterer with the Rayleigh-type source of scattering. The
proposed here mechanism leads to the strong  decrease of
$l(\omega )$ (plateau-like regime) at higher frequencies. As the
result, the plateau region of quasilattice $\kappa $ exists at
higher temperatures than constant $\kappa $ of amorphous
SiO$_{2}$ from\cite{and}. Besides, the absolute value of
quasilattice $\kappa $ in $T$-independent region exceeds the
corresponding value for SiO$_{2}$ (See Fig.2) with the same
fixed parameters related to biaxial WDD. This result is in
agremeent with that obtained in\cite{chern1}.

In Fig.1 the experimental data are well fitted with our
theoretical curves at highest temperatures (above $T\approx
50$K) where  $\kappa (T)$ slightly increases. In our
calculations this rise of $\kappa $ is the result of $l_{min}$
introduction ( see details above in text). However, the physical
reason of this increase is, evidently, in opening of a new
heat-carrying channel e.g. localized modes (fractons) hopping
mechanism\cite{janot}. Interestingly enough, in our model as was
mentioned above  the average value of the dipole separation $2L$
has the same meaning as $l_{min}$ ($\sim 20\AA$). Thus, $2L$  or
in other words an average grain boundary size  between two
adjoining icosahedral clusters can serve as a parameter of
phonon localization. An intercluster size as the possible origin
of localization of the modes  has been considered
in\cite{boiss}.

In conclusion, a new theoretical approach was proposed to fit
the experimentally observed thermal conductivity for icosahedral
$i$-AlPdMn quasicrystals. This mechanism of phonon scattering
implies that cluster interfaces in this class of compounds are
responsible for the phonon scattering at very low temperatures.
At low and intermediate temperatures a good agreement between
our model and experimental data was reached by a combination of
cluster interface and quasiumklapp scattering mechanisms.

Author thanks V.A. Osipov for helpful discussions.

\end{document}